\documentclass[aip,graphicx,twocolumn,10pt,apl]{revtex4}
\usepackage{graphicx} \usepackage{epstopdf}

\usepackage{afterpage}
\usepackage{times}

\begin{document}
\title{Large electric field effects on the resistance of La$_{0.67}$Ca$_{0.33}$MnO$_3$ microstructures}
\author{C. Beekman, I. Komissarov and J. Aarts}
\affiliation{Kamerlingh Onnes Laboratorium, Leiden University, The
Netherlands}
 \email[E-mail:]{aarts@physics.leidenuniv.nl}

\date{\today}
\begin{abstract}\noindent
We investigate electric field effects in thin film microbridges of
La$_{0.7}$Ca$_{0.3}$MnO$_3$ with the focus on the regime of metal-insulator
transition. A mechanically milled SrTiO$_3$ substrate is used as a backgate
dielectric. Inside the metal-insulator transition we find a strong unipolar
field-induced reduction in resistance, as well as a suppression of the
nonlinear features in the I-V curves we observed earlier. We associate the
observed effects with a phase separated state in which metallic regions coexist
with short range correlated polaron regions. When the glassy polaron phase has
fully developed, and closes off the microbridge, the field effects disappear
leaving the strongly nonlinear behavior of the transport current unaltered.
\end{abstract}
\pacs {} \maketitle
\begin{figure}[t]
\includegraphics[width=7.2cm,height=4.8cm,angle=0]{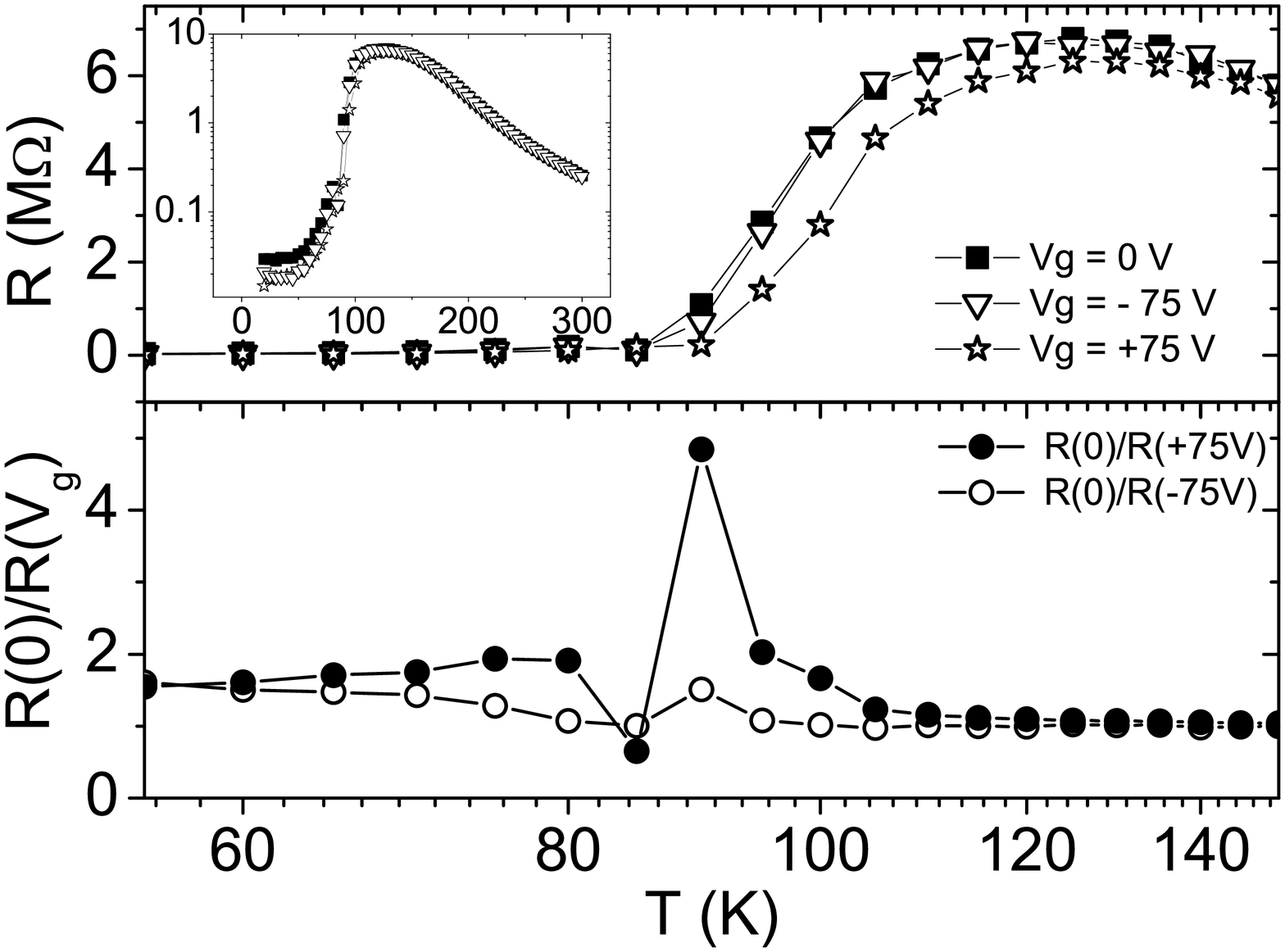}
\centering\caption{(Top) R(T) behavior of a 5 $\mu$m x 10 nm L7CMO bridge
measured at I = 0.1 $\mu$A for three applied gate voltages: V$_g$ = 0, +/-75 V
(Inset: full R vs. T curves). (Bottom) ratio between R(0) and R (V$_g$) (closed
circles: V$_g$ = + 75 V; open circles: V$_g$ = - 75 V). }\label{gate2}
\end{figure}

\begin{figure}[t]
\includegraphics[width=7.2cm,height=4.8cm,angle=0]{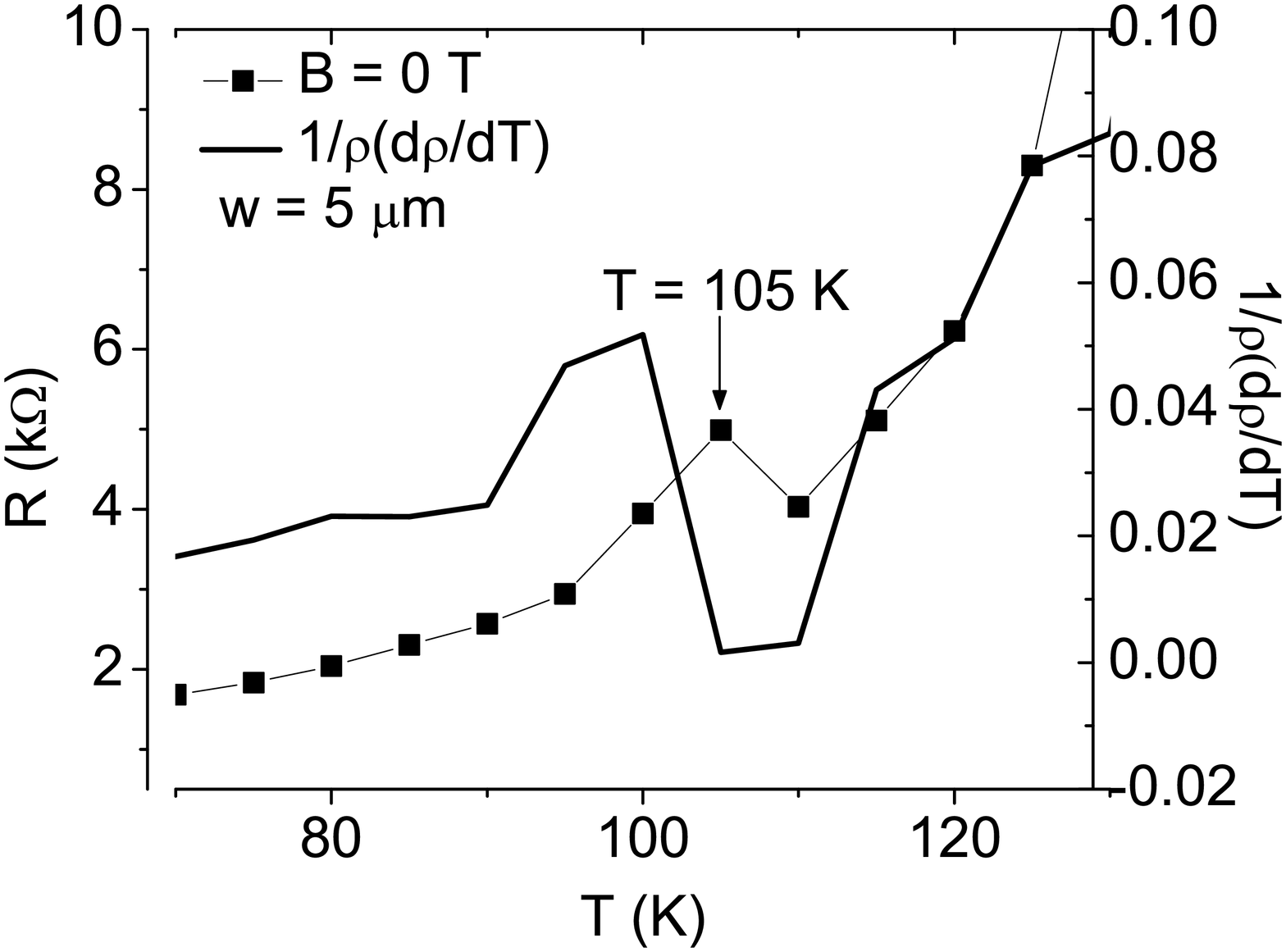}
\centering\caption{R(T) behavior of a 20 nm thick L7CMO microbridge (w = 5
$\mu$m ) between T =50 - 135 K. Right axis: 1/$\rho$(d$\rho$/dT) (solid line).
The arrow indicates the transition in the STO. Inset: full R(T)
curve.}\label{61b}
\end{figure}

 \vspace{-0.5cm} \noindent
 Conventional field effect transistors (FET) consist of semiconductor channels in
which the carrier density is modulated by an electric field applied through a
gate on top or below the channel. The dimensions of these semiconductor
microstructures are reaching their intrinsic physical limit. Further increase
of circuit density requires the consideration of different materials for both
the gate dielectric and the channel. Promising are correlated electron systems
based on Mott insulators, such as high T$_C$ superconductors\cite{ahn6b} and
Colossal Magnetoresistance (CMR) manganites \cite{ahn6c}. The abundant amount
of potential carriers (d-electron) and the ability to control the bandgap
could result in novel FET-type~devices. \\
\indent Doped manganites such as La$_{0.7}$Ca$_{0.3}$MnO$_3$ (L7CMO), obtained
by hole doping the antiferromagnetic insulating parent compound LaMnO$_3$, show
a large variety in physical properties \cite{tokura1,dagotto}. The Ca-doping
introduces mixed Mn-valence, with both Mn$^{3+}$ present, which is Jahn-Teller
(JT) active, and Mn$^{4+}$, which is not. This leads to competing interactions,
trapping of electrons in JT distortions (polarons) and itinerancy of the
electrons in the Double Exchange (DE) mechanism\cite{zener} when spins become
polarized. Depending on the doping, this leads to a metal-to-insulator
transition (from low to high temperature), at a characteristic temperature
T$_{MI}$. \\
Epitaxial all perovskite FET-devices already have been under investigation. For
example a 3 nm La$_{0.8}$Sr$_{0.2}$MnO$_3$ film (screening length less than 1
nm\cite{hong6},) combined with a ferroelectric material (PbZr$_{0.2}$
Ti$_{0.8}$O$_3$ , PZT) as the gate dielectric, shows $E$-field induced
modulation of T$_{MI}$ and the magnetoresistance. These effects \cite{hong6,
pallechi6b} are \emph{bipolar} (i.e. the sign of the resistance change is
opposite for opposite signs of the applied gate voltage) and attributed to the
modulation of the charge carrier density in the manganite, in other words a
straightforward modulation of the doping effect. In a different investigation,
large in-plane resistance variations ($\sim$ 76 $\%$, bipolar) were observed in
a 50 nm thick L7CMO films\cite{wu6} with the PZT-gate geometry, and smaller
effects for devices deposited on an a SrTiO$_3$ (STO) film as the backgate
dielectric. Since the thickness of the film is much larger than the $E$-field
screening length, the effects were attributed to the presence of a phase
separated state, the more so since there was clear asymmetry for the two signs
of the gate voltage. For phase separation, fully \emph{unipolar} field effects
were also observed, for example in La$_{0.8}$Ca$_{0.2}$MnO$_3$ thin films on
mechanically thinned STO substrates \cite{eblen6}. \\
\indent What has been little investigated is the effect of an applied $E$-field
on a microbridge, where the channel width may be comparable to the length
scales associated with the phase separation phenomenon. Here we report
$E$-field effects on strained L7CMO microbridges using the STO substrate as a
backgate dielectric. Our microbridges show a metal-insulator transition at
T$_{MI}$ and the well-known CMR effect, typical for strained thin films. They
also show non-linear current $I$ - voltage $V$ characteristics in the range of
the MI-transition, on which we reported before \cite{beekman}. We find strong
\emph{unipolar} field effects in the onset of the transition, which we
associate with the occurrence of a phase separated state, in which metallic
regions coexist with short-range correlated polaron regions\cite{lynn}. As the
system is warmed through the M-I transition the field effects disappear when
the more or less homogeneous correlated polaron (glass) phase is fully
developed. \\
\begin{figure}[b]
\includegraphics[width=8cm,height=5.5cm,angle=0]{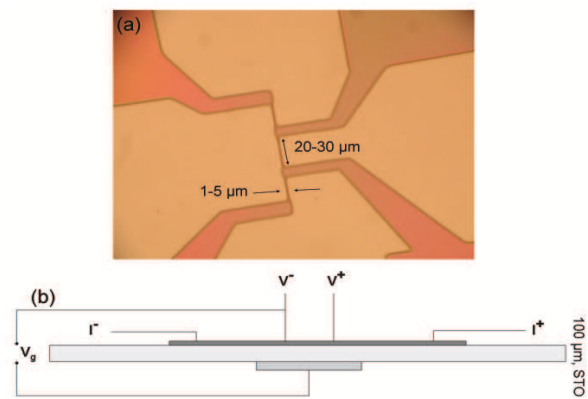}
\centering\caption{a) The microbridge patterned into the L7CMO thin film.
Dimensions: width: 5 $\mu$m and the distance between the voltage contacts is 30
$\mu$m. b) FET device geometry. }\label{FET}
\end{figure}
\indent The L7CMO films were grown by DC-sputtering in an oxygen atmosphere of
3 mbar and at a growth temperature of 840 $^{\circ}$C. The films are patterned
into microbridges\cite{beekman2} using electron-beam lithography and Ar-etching
(see Fig.\ref{FET}b). The STO substrate was subsequently mechanically milled
down to 100 $\mu$m and used as a gate dielectric. Measurements before and after
the milling show that the bridges remain undamaged during this process. The
geometry of the measured devices is shown in Fig.\ref{FET}a. The gate voltage
V$_g$ is applied between the back of the STO substrate (through a silver paint
contact) and one of the voltage contacts of the microbridge. We measured I-V
curves as function of temperature and in high magnetic fields using a Physical
Properties Measurement System (Quantum Design) for temperature and magnetic
field control (T = 20 - 300 K; H$_a$ = 0 - 9 T). We found the leak currents
through the gate to be negligible compared to the currents used for the I-V
measurements. \\
\indent I-V characteristics were measured in applied electric fields up to
1x10$^6$ V/m for microbridges patterned in 10 nm thick L7CMO films grown on STO
substrates.
\begin{figure}[t]
\includegraphics[width=6.5cm,height=4cm,angle=0]{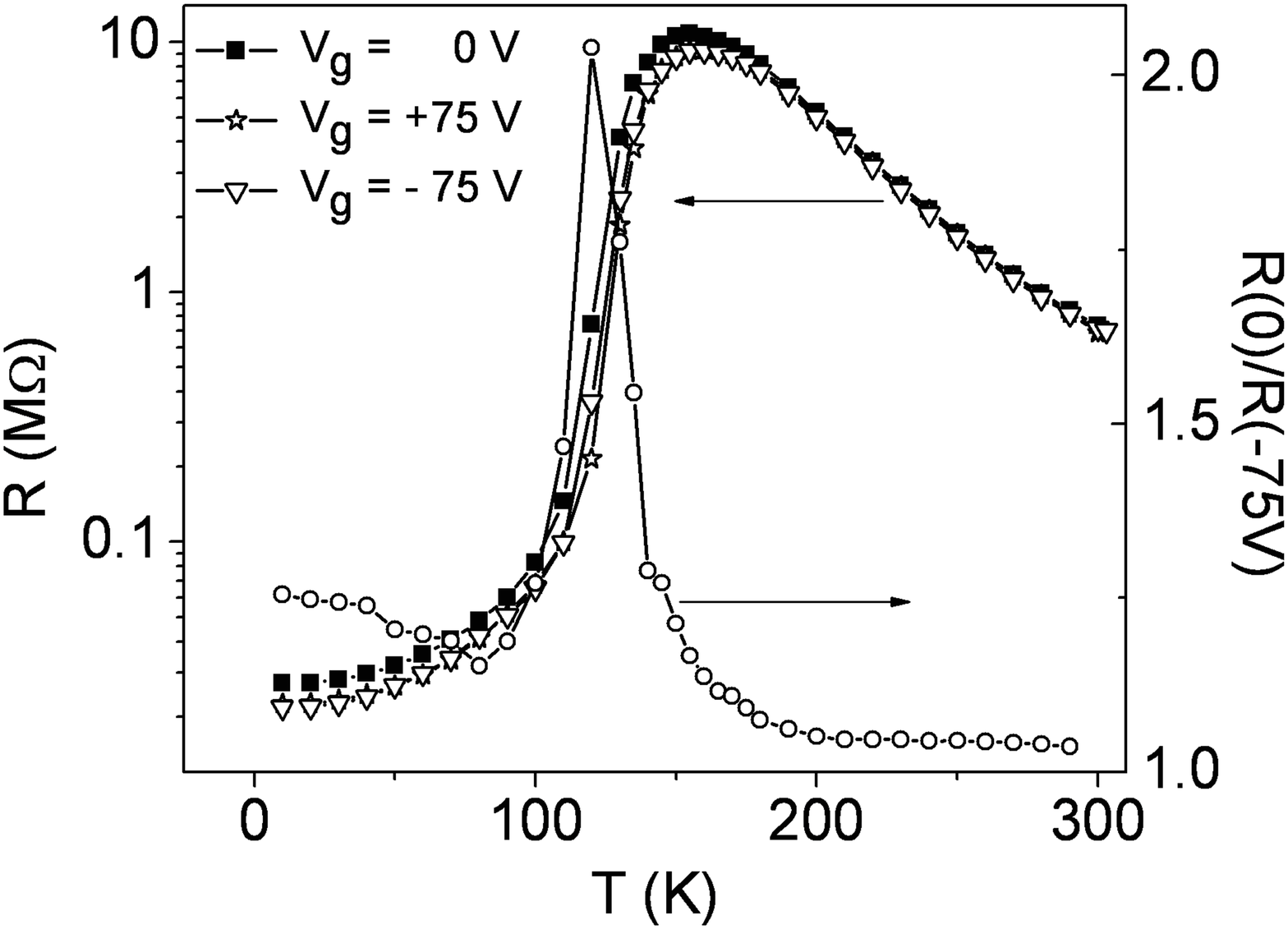}
\centering\caption{R(T) behavior (left axis) of a 500 x 10 nm$^2$ L7CMO bridge
determined from I-V curves at I around 0.1 $\mu$A for three applied gate
voltages: V$_g$ = 0, +/-75 V. On the right axis the ratio between R (0 V) and R
(-75 V) is shown. }\label{gate1}
\end{figure}
In Fig.\ref{gate1} we show the resistance behavior of the 0.5 $\mu$m bridge in
zero field and upon application of +/- 75 V (i.e. 7.5x10$^5$ V/m). In zero
field we find typical resistance vs. temperature behavior for strained L7CMO
films\cite{beekman3}. Furthermore, we observe a strong $E-$field effect in the
transition around T = 120 K (T$_{MI}$ $\approx$ 160 K). The resistance is
reduced by a factor of 2 when V$_g$ = -75 V is applied. Application of a
positive gate voltage shows a smaller reduction, the effect is \emph{unipolar}.
It is important to note is that the position of T$_{MI}$ remains unchanged when
gate voltages are applied, and also that the effect is sharply peaking in the
regime of the MI transition and not beyond, quite different from the
obswrvations by Wu {\it et al.}\cite{wu6}
Repeating this measurement on a 5 $\mu$m bridge (Fig.\ref{gate2}) we observe a
similar effect. In this case the effect is also sharply peaked, now at T = 90 K
(T$_{MI}$ $\approx$ 120 K), and the resistance is reduced by a factor of 5 upon
application of V$_g$ = +75 V, with a somewhat larger asymmetry (compared to the
0.5 $\mu$m microbridge) between gate voltages of opposite sign. In both
microbridges we also observe a (smaller) reduction in resistance at low
temperatures. Both signs of the gate voltage result in similar resistance
changes. In this temperature regime, once an $E$-field has been applied the
microbridge appears to be irreversibly changed. For the microbridge in
Fig.\ref{gate2} the initial low temperature (higher resistivity) state was not
recovered after gating. However, repetition of the same experiment but with
application of 100~V did show very similar behavior around the transition
albeit with slightly reduced magnitude for the field induced resistance drop
(data not shown).\\
\indent Next we turn to the I-V curves measured on the 5~$\mu$m sample. In
accordance with previous findings\cite{beekman} they are linear for most
temperatures but show strong nonlinear behavior in the steep part of the
transition. Here we investigate the influence of the applied electric field on
these nonlinearities.
\begin{figure}[b]
\includegraphics[width=7.5cm,height=5cm,angle=0]{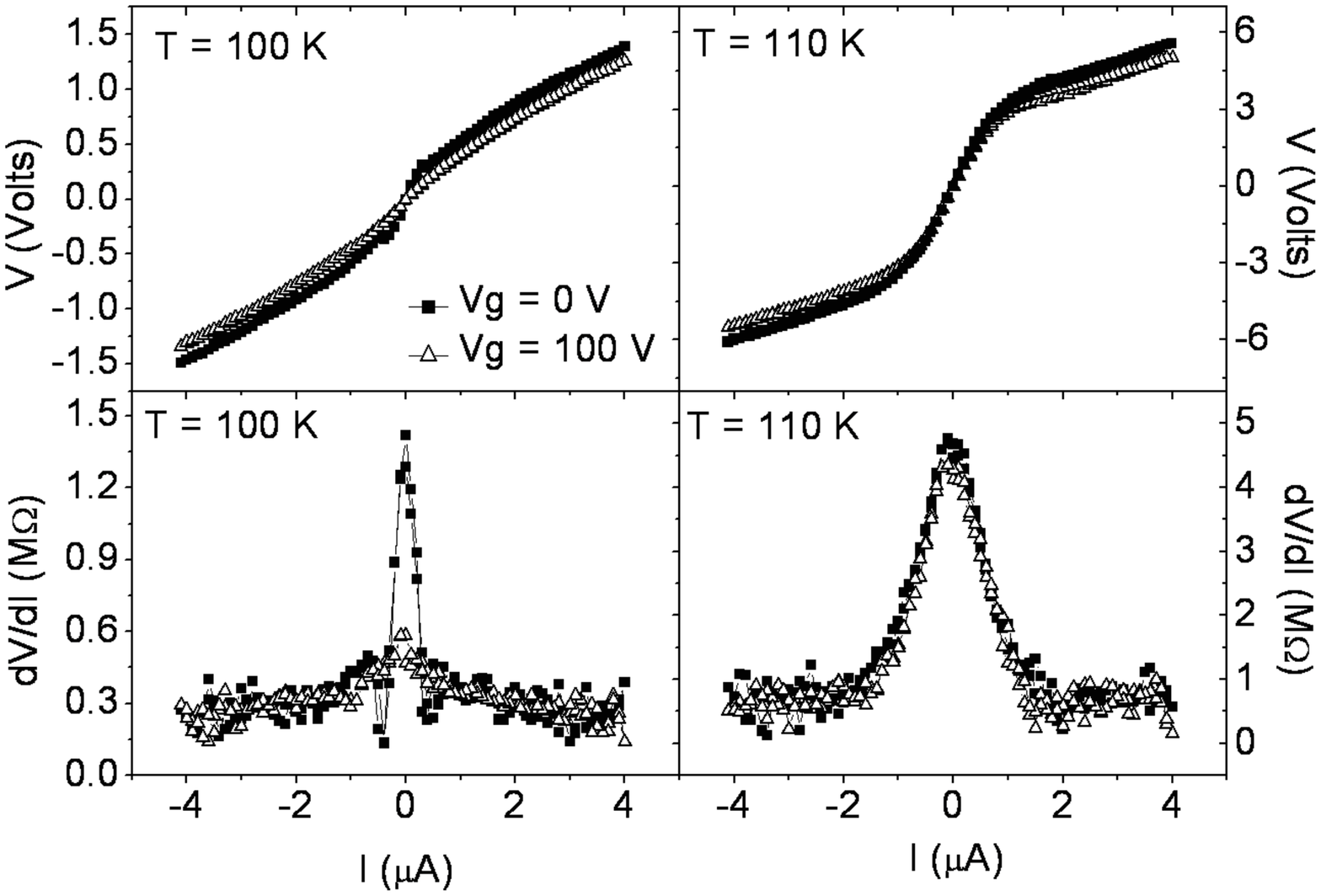}
\centering\caption{(Top) I-V curves for 5 $\mu$m bridge at T = 100 K and T =
110 K. (Bottom) corresponding numerical derivatives.  V$_g$ = 0 V (closed squares) and V$_g$ =
+100 V (open triangles).}\label{gate5}
\end{figure}
Fig.\ref{gate5} shows two I-V curves (5 $\mu$m bridge, T = 100 K and T = 110 K)
and their corresponding (numerical) derivatives for V$_g$ = 0 V and V$_g$ =
+100 V. At 100~K the nonlinearities which we associated with the formation of a
homogeneous glassy polaron phase start to appear, with a full width of the peak
in dV/dI of 1 $\mu$A. From the data it becomes clear that at this temperature
the nonlinearity is suppressed upon application of an $E$-field. At T = 110 K,
where the nonlinearity has developed strongly (peak width: 4 $\mu$A), the
strength and shape of the peak remains fully unaltered when an $E$-field is
applied.\\
\indent One concern with respect to our observations might be that they are
connected the cubic-to-tetragonal phase transition in STO around 105~K. A small
effect was actually found to exist in thin films of L7CMO on STO, not in the
resistivity, but in the temperature coefficient (TC) 1/$\rho$(d$\rho$/dT) (with
$\rho$ the specific resistance), which showed a variation of 0.5\% in a 9~nm
film\cite{egilmez6}. In that film T$_p$ was at 160~K, and the TC variation was
observed in the flat metallic part of the resistance. Our thin films have lower
T$_p$, which we have argued is the effect of homogeneous strain\cite{beekman3},
and a 1\% variation is not visible in the strong decrease in $\rho$ below the
MI transition. We could detect a variation  in the TC in the microbridge of a
20~nm film where T$_p \approx$ 160~K. As shown in Fig.\ref{61b} we even find a
variation in $R(T)$, which was not yet reported before. This is clearly due to
the larger film homogeneity in the small structure. Another point to be made is
that there is hardly any change in the dielectric constant of STO at the phase
transition\cite{salje91}. Our $E$-field effects appear to be intrinsic features
of the L7CMO microstructures. \\
\indent We explain the observations in light of our previous report on the
formation of a polaron glass phase in L7CMO microbridges as it is warmed
through the M-I transition\cite{beekman}. The \emph{unipolar} nature of the
observed effects indicate that they are to be attributed to a state in which
metallic and insulating regions coexist, and are influenced by the $E$-field.
The insulating regions are formed by short range correlated polarons, as
precursor to the polaronic state at high temperature. The applied field changes
the relative volume fraction of the coexisting phases by accumulating charge at
the interfaces between them, which can result in the dielectric breakdown of
the inhomogeneities. Important to note is that the maximum in the effect occurs
in the onset of the transition when the nonlinearities just start to appear in
the $I$-$V$ curves. When the nonlinear effect has fully developed, the electric
field effect disappears. Apparently, when the glassy polaron phase becomes
homogeneous and closes off the bridge, the $E$-field cannot break it down. We
note that the effect in the transition is unipolar but asymmetric. It is
possible that doping still plays a role and that asymmetry in hole and electron
modulation of inhomogeneities in the microbridge lead to the observed
asymmetric behavior. \\
We also note that the effects in the transition are different from those at low
temperatures, where it is irreversible or hysteretic, indicating that this
regime is also not inhomogeneous, and that the collapse and rebuilding of  the
insulating regions is not well controlled \cite{dong6}. The microbridge on the
other hand appears to at least mostly relax back to its initial state since
remeasuring the gate effect (days later) leads to similar but somewhat smaller
effects in the transition. \\
\indent In conclusion, going to micron-sized structures reveals a strong
response of La$_{0.7}$Ca$_{0.3}$MnO$_3$ thin films to an applied electric field
which has not been reported before, and which is clearly tied to the
percolating behavior of the conductance in the bridge, which takes place on the
scale of the width of the bridge.  \\
\indent We are grateful for discussions with J. Zaanen. This work was part of
the research program of the Stichting F.O.M., which is financially supported by
NWO.
\end{document}